\documentclass[11pt,twoside]{article}
\usepackage{macro-fsut-eng}
\usepackage{graphicx}
\usepackage{subfigure}

\usepackage[T1]{fontenc} % Computer Modern (CM) fonts

\usepackage{latexsym}
\usepackage{verbatim}

\begin{document}

\vskip 1.0cm
\markboth{H.~Velten and S.~Calogero}{Exploring cosmological matter diffusion coefficients}
\pagestyle{myheadings}

\vspace*{0.5cm}
\title{Exploring non-linear cosmological matter diffusion coefficients}

\author{Hermano~Velten$^1$ and Simone Calogero$^2$}
\affil{$^1$Departamento de F\'isica, Universidade Federal do Esp\'{\i}rito Santo (UFES), Av. Fernando Ferrari, Campus Goiabeiras, Vit\'oria, Brazil\\
$^2$Mathematical Sciences, Chalmers University of Technology, Gothenburg University,S-412~96 Gothenburg}

\begin{abstract} 
Since microscopic velocity diffusion can be incorporated into general relativity in a consistent way, we study cosmological background solutions when the diffusion phenomena takes place in an expanding universe. Our focus here relies on the nature of the diffusion coefficient $\sigma$ which measures the magnitude of such transport phenomena. We test dynamics where $\sigma$ has a phenomenological dependence on the scale factor, the matter density, the dark energy and the expansion rate.
\end{abstract}

\section{Introduction}

For most of the universe lifetime its dynamics can be approximated by a simple expanding dust matter dominated sphere. The hot radiative (Big Bang) primordial universe cools down quickly until the radiation energy drops to the same level as the matter energy density. This happens very soon, when the universe is only $\sim$ 50 Kyrs old. During the following $10$ Gyrs the total universe cosmic energy budget is well approximated by a pressureless matter fluid. This is the matter dominated epoch where most of the main astrophysical effects take place, such as the formation of stars, galaxies and clusters of galaxies.
The matter component can be divided into two distinct contributions: the first one is the expected baryonic sector which contains the known heavy particles of the standard particle model. The second contribution comes from an unknown component called dark matter which is at least five times more abundant than the baryonic matter and is the building block of any successful cosmological theory. The matter domination era is a necessary stage for the formation of structures, but it ends when the universe is $\sim 10$ Gyrs old. From this moment on, another form of energy, called dark energy, accelerates the background expansion slowing down the agglomeration rate. The nature of the dark energy is also still unknown. The simplest explanation for this effect relies on the existence of a cosmological constant $\Lambda$. However, one could admit different descriptions for the dark energy phenomena, like scalar fields, which may (Amendola 2000; Zimdahl et al 2003; Dalal et al 2011; Castro et al 2012) or may not interact with the other cosmic components.

In the standard model described above the matter dynamics is therefore described by the relativistic Euler equation $\nabla^\mu T_{\mu\nu}=0$ on the the matter fluid energy-momentum tensor. In particular, the fluid feels only indirectly (via the gravitational potential) the presence of other components, e.g., photons, neutrinos and dark energy. %This approach is based on the assumption that dark matter behaves as a (presureless) perfect fluid described by . Assuming that the 
If we assume fluid particles undergoing velocity diffusion in a background medium, it was shown in~ (Calogero 2011; Calogero 2012) that the matter dynamics can be described by the equations
\begin{equation}\label{cons}
\nabla_\mu J^\mu=0,\quad
\nabla_\mu T^{\mu\nu}=\sigma J^\mu.
\end{equation}
The first equation guarantees the standard conservation law for the particles current density $J^\mu$. The quantity $\sigma$ in the second equation is the (positive) diffusion coefficient, which measures the energy transferred to the fluid particles by the diffusion forces per unit of time\footnote{We take the opportunity to point out a little missprint in our article~(Calogero \& Velten, 2013). In the paragraph after eq. (2.3), the sentence ``The value $3\sigma$ measures the energy..." should read ``The value $\sigma$ measures the energy..." }. 
So far only the case of a constant $\sigma$ has been considered in the literature, see e.g.~(Calogero \& Velten,  2013; Shogin et al 2013), but here the possibility that $\sigma$ varies through space-time will be considered. 

%However, the main consequence of admiting the diffusion mechanism to take place in a covariant theory is reflected in the equation
Since the second relation in~(\ref{cons}) states that, in the presence of diffusion, the matter energy-momentum tensor is not a divergence free quantity, and having in mind Bianchi's identities, it is clear that the space-time geometry cannot be determined by the standard Einstein field equations of general relativity.  
The inconsistency with the Bianchi identities can be circumvented by adding a cosmological scalar field $\phi$ to the Einstein equation, which thereby becomes
\begin{equation}\label{EinsteinEq}
R_{\mu\nu}-\frac{1}{2}g_{\mu\nu}R+\phi g_{\mu\nu}=T_{\mu\nu},
\end{equation} 
where we use physical units such that $8\pi G=c=1$.
The scalar field $\phi$ plays the role of the background medium in which diffusion takes place. 
Taking the divergence $\nabla^\mu$ of both sides of eq.~(\ref{EinsteinEq}), we obtain that $\phi$ obeys
\begin{equation}\label{phieqfluid}
\nabla_\mu\phi=\sigma n J_\mu.
\end{equation}
% The value $3\sigma$ measures the energy transferred from the scalar field to the matter per unit of time due to diffusion since we have used units   
%Taking a divergence of both sides of~(\ref{phieqfluid}) we obtain that $\phi$ satisfies the wave equation
%\begin{equation}\label{wave}
%\Box\phi=J^\mu\nabla_\mu\sigma.
%\end{equation}
%Note that for $\sigma\equiv 0$ we recover the standard cosmological constant scenario. It is therefore natural to identify $\phi$ with the dark energy component of the universe.

In order to avoid the need to introduce a new evolution equation for $\sigma$, and at the same time to ensure that the value of $\sigma$ is coordinates-independent, we assume that $\sigma=f(s)$, where $s$ is a scalar invariant quantity constructed from $\phi, g_{\mu\nu}, J^\mu$ and $T_{\mu\nu}$. The simplest choices for the scalar invariant $s$ are
\begin{equation}\label{invariants}
s_1=-J^\mu J_\mu, \quad   s_2=g^{\mu\nu}T_{\mu\nu}, \quad s_3=\phi.
\end{equation}
 
In the next section we present the basic equations for a viable cosmological model based on the diffusion theory outlined above. This model extends the one studied in (Calogero \& Velten,  2013) by considering a time dependent diffusion coefficient $\sigma$. 
%. However, while our previous investigation was limited to the case where the matter diffusion coefficient is constant  during the entire universe expansion, in this contribution we briefly explore the consequences of assuming 

%We present our parameterizations for $\sigma$ in section 3. In the same section we present our results. We will be mostly concerned with the behavior of background quantities, such as the normalized density parameters and the Hubble expansion rate. 
%and the deceleration parameter. We conclude in the final section with a few comments and suggestions on possible generalizations of this work.

\section{Cosmological model with variable matter diffusion}

A viable cosmological model in which dark matter undergoes microscopic velocity diffusion into a dark energy solvent field $\phi$ has been developed in (Calogero 2012; Calogero \& Velten,  2013). This model is obtained from the general diffusion theory described in the Introduction under the following assumptions: (i) the matter content is described by a pressureless fluid, i.e., the energy-momentum tensor $T_{\mu\nu}$ and the current density $J^\mu$ are given by $T_{\mu\nu}=\rho\, u_\mu u_\nu$ and $J^\mu=n u^\mu$, where $\rho$ is the energy density, $n$ the particles number density and $u^\mu$ the four-velocity field of the dust fluid; (ii) the universe is spatially homogeneous, isotropic and flat and so in particular the space-time metric can be written in the form
\[
ds^2=-dt^2+a(t)^2(dx^2+dy^2+dz^2),\quad a_0=1,
\]
where a subscript 0 indicates the evaluation at time $t=0$; (iii) the diffusion coefficient $\sigma$ is a positive constant. The resulting cosmological model has been called $\phi$CDM model in (Calogero \& Velten,  2013) and is described by the following system on the standard normalized energy densities
\begin{eqnarray}
&&\frac{d\Omega_m(z)}{dz}-\frac{3\Omega_m(z)}{1+z}=-\tilde{\sigma}\frac{(1+z)^2}{E(z)},\label{dimensionlesseq1new}\\
&&\frac{d\Omega_\phi(z)}{dz}=\tilde{\sigma}\frac{(1+z)^2}{E(z)},\label{dimensionlesseq2new}\\
&&\frac{H(z)}{H_0}=E(z)=\sqrt{\Omega_m(z)+\Omega_\phi(z)},\label{dimensionlesseq3new}
\end{eqnarray} 
with 
\begin{equation}
z=a^{-1}-1,\quad \tilde{\sigma}=\frac{\sigma n_0}{3H^3_0}.
\end{equation}
For $\tilde{\sigma}=0$ the $\phi$ field remains constant in time and the solution is given by the $\Lambda$CDM model:
\[
\Omega^{(0)}_m(z)=\Omega^{(0)}_{m0}(1+z)^3,\quad \Omega_\phi^{(0)}(z)=\Omega_{\phi0}^{(0)}=1-\Omega^{(0)}_{m0}.
\]
Equations (\ref{dimensionlesseq1new}) and (\ref{dimensionlesseq2new}) denote a coupled system where energy flows from the matter to the dark energy field. The direction of the flux is due to the fact that $\tilde{\sigma} > 0$. Interacting models play an important role to alleviate the cosmic coincidence problem, i.e., the fact that only today the dark matter and dark energy densities are of the same order of magnitude. Usually the interaction term in the right hand side of equations (\ref{dimensionlesseq1new}) and (\ref{dimensionlesseq2new}) is incorporated in a {\it ad hoc} way. Therefore, the diffusion mechanism appears as a genuine physical mechanism responsible for the interaction on the dark sector.

In the rest of the paper we assume that $\tilde{\sigma}$ in the equations above is time-dependent. 
We will employ the following phenomenological choices

\[
\tilde{\sigma}_{(n)} \equiv \tilde{\sigma}_0 a^k,\quad \tilde{\sigma}_{(\rho)} \equiv \tilde{\sigma}_0 \left(\frac{\Omega_m}{\Omega_{m0}}\right)^{\lambda},\quad \tilde{\sigma}_{(\phi)}\equiv\tilde{\sigma}_0 \left(\frac{\Omega_\phi}{\Omega_{\phi0}}\right)^{\delta}, \quad \tilde{\sigma}_{(H)} \equiv \tilde{\sigma}_0 \left(\frac{H}{H_0}\right)^{h}.
\]
We remark that, since for the model under discussion the scalar invariants $s_1,s_2$ in (\ref{invariants}) are given by $s_1=n_0(1+z)^3$, and $s_2=-\rho=H_0\Omega_m(z)$, the choices $\tilde{\sigma}_{(n)}$, $\tilde{\sigma}_{(\rho)}$ and $\sigma_{(\phi)}$ correspond respectively to a diffusion coefficient that is a power of the scalar invariants $s_1=n^2,s_2=-\rho,s_3=\phi$ to which~(\ref{invariants}) reduce in the dust fluid case.

\section{Cosmological background dynamics}
Let us investigate now how the different options for the coefficient $\sigma$ affect the background dynamics of the cosmological model. 

In our analysis we will fix the reference values $\Omega_{m0}=0.3$ $(\Omega_{\Lambda}=\Omega_{\phi 0}=0.7)$ and $H_0=70 km \,s^{-1}Mpc^{-1}$. Moreover the today magnitude of the diffusion coefficient will be fixed at $\tilde{\sigma}_0=0.1$. Although background observational data can be described by this value, the structure formation process is severely affected by the diffusion mechanism. The analysis using the matter power spectrum data imposes the upper bound $\tilde{\sigma}_0< 0.01$ (Calogero \& Velten,  2013). However, we will keep the reference value $\tilde{\sigma}_0 = 0.1$ as a guide since here we are mostly concerned with the background expansion. Indeed, depending on the value of $k$, $\lambda$, $\delta$ and $h$, the resulting diffusive dynamics becomes closer to the $\Lambda$CDM model, thus allowing for larger values of $\tilde{\sigma}_0$.

\begin{figure}
\begin{center}
\subfigure[$\tilde{\sigma}_{(n)}$]{\includegraphics[width=0.45\textwidth]{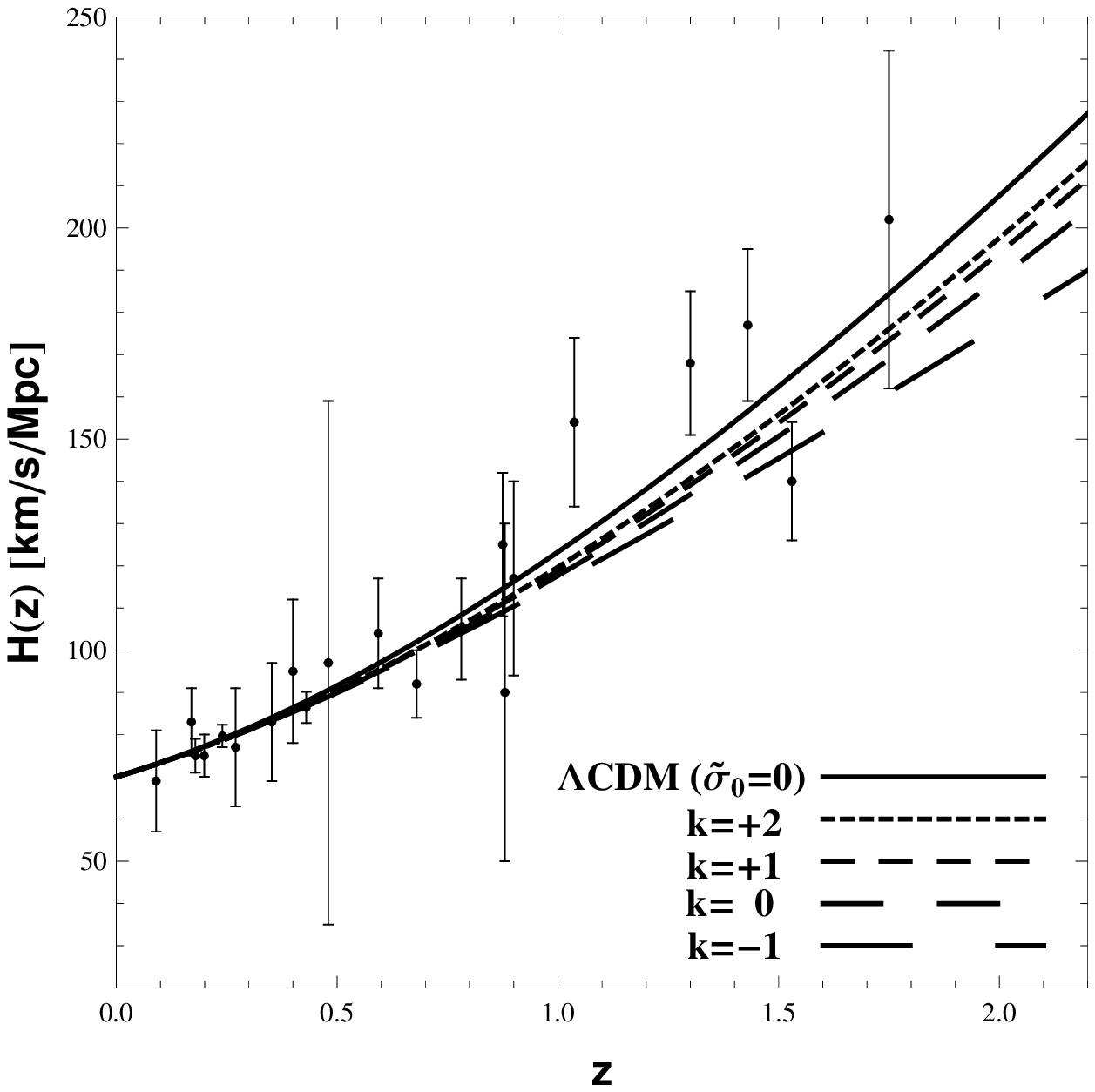}}\quad
\subfigure[$\tilde{\sigma}_{(\rho)}$]{\includegraphics[width=0.45\textwidth]{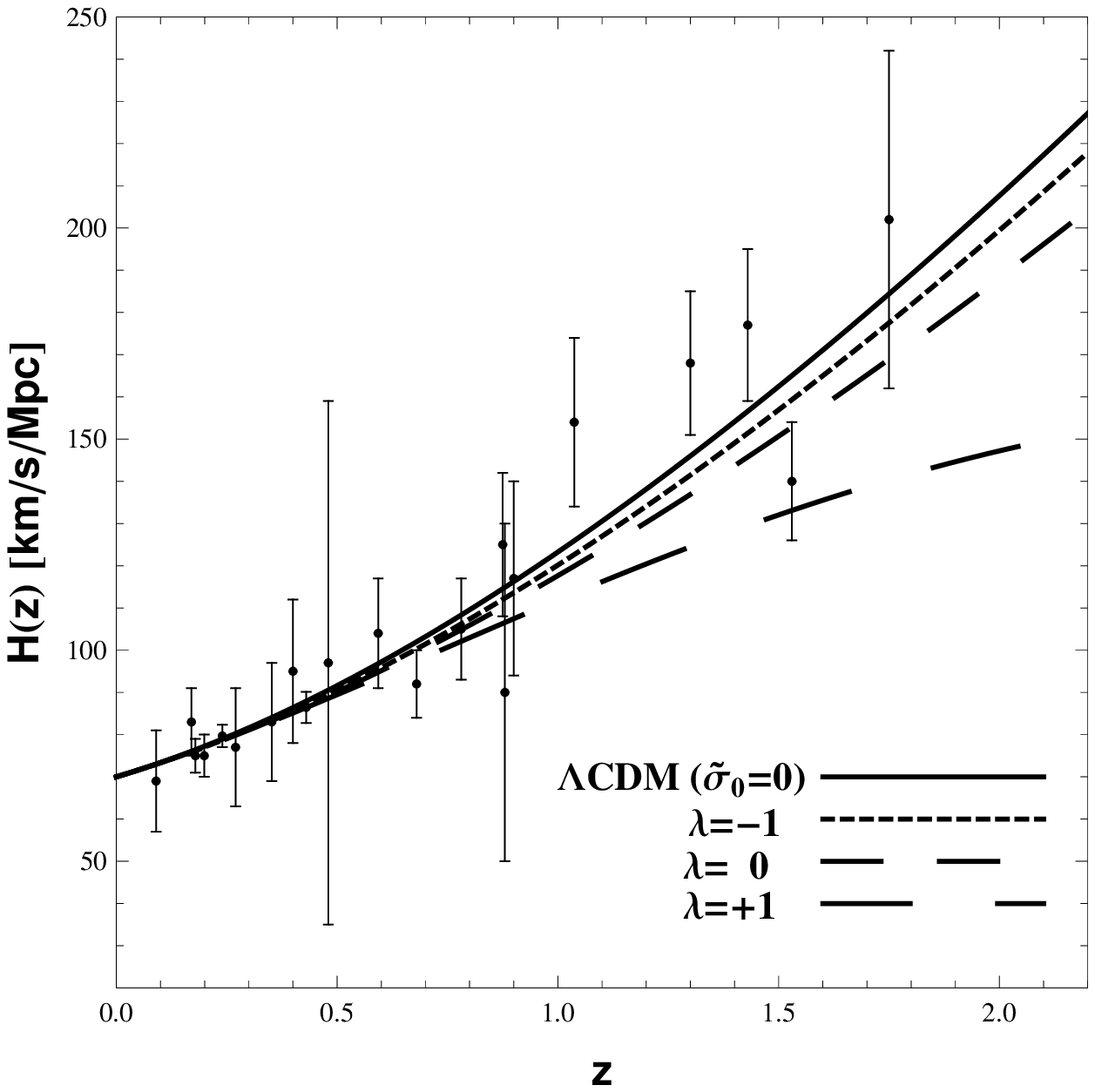}}\\
\subfigure[$\tilde{\sigma}_{(\phi)}$]{\includegraphics[width=0.45\textwidth]{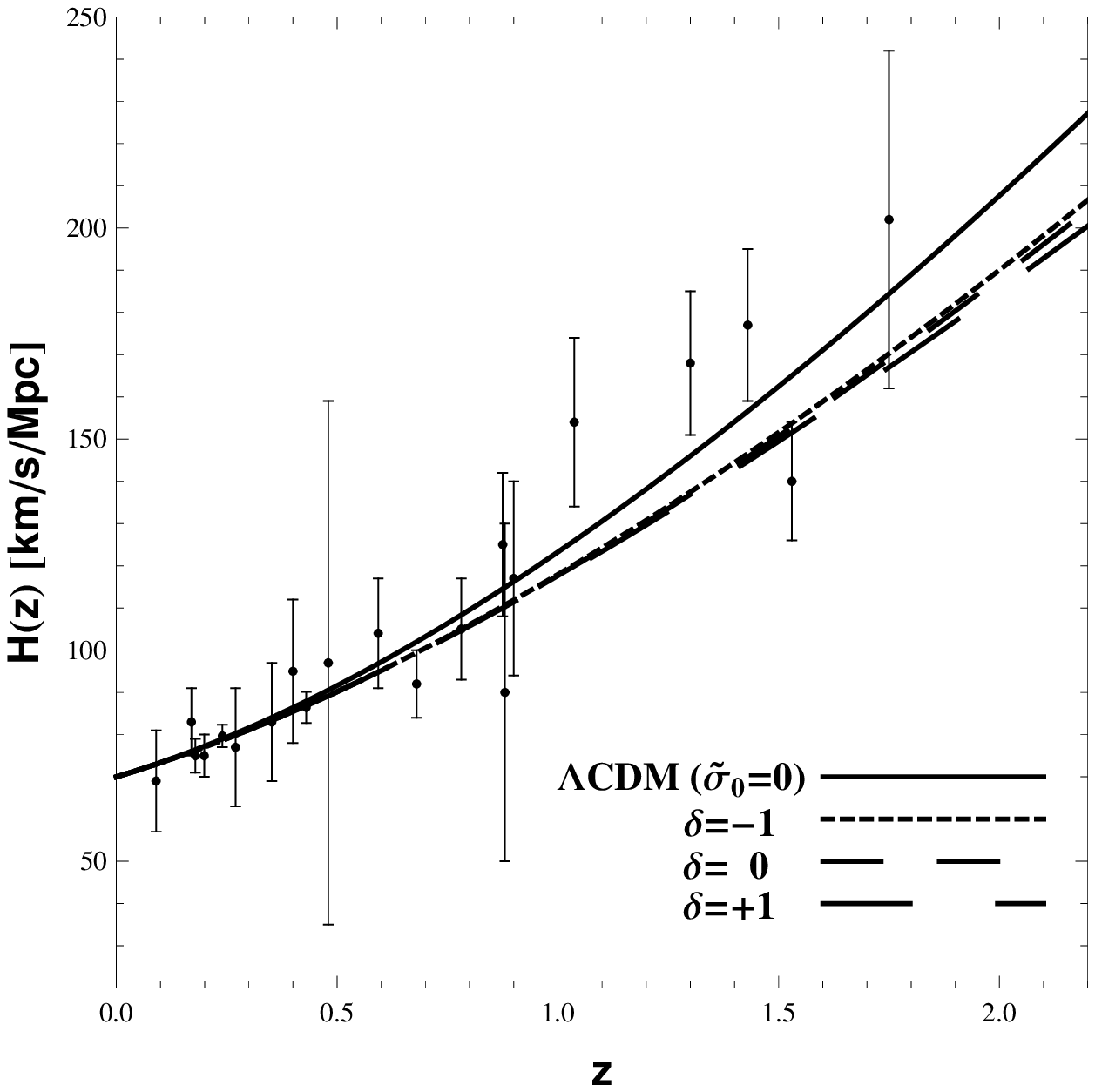}}\quad
\subfigure[$\tilde{\sigma}_{(H)}$]{\includegraphics[width=0.45\textwidth]{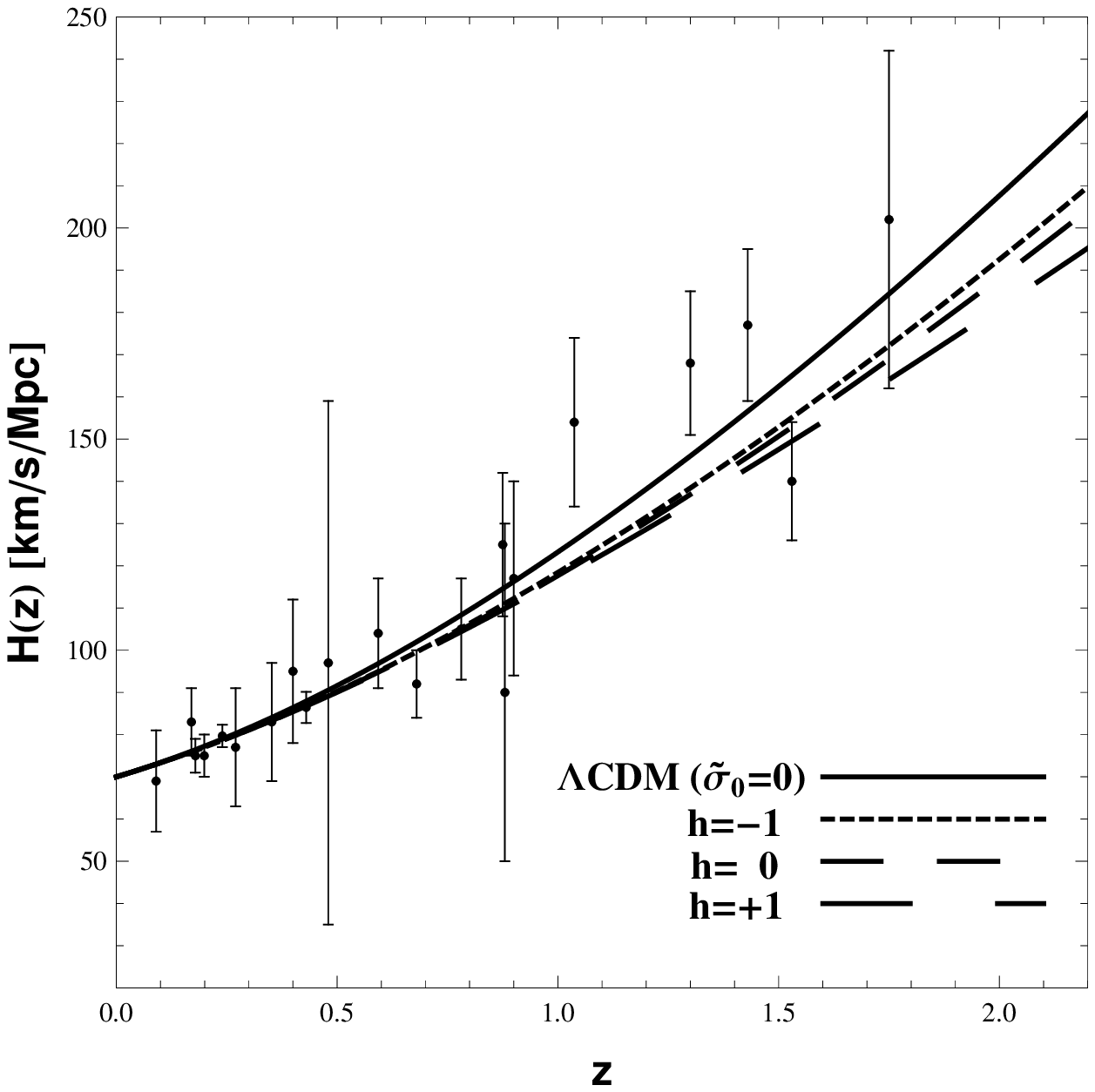}}
\caption{Evolution of the Hubble function for the various choices of time dependent diffusion coefficient.}
\label{fig1}
\end{center}
\end{figure}

The results of our analysis are contained in Figs. 1 and 2, where we plot respectively the Hubble function and the fractionary densities corresponding to the different choices of the time dependent diffusion coefficient $\sigma$. In each plot the dynamical quantities for the $\Lambda$CDM model are shown with solid lines. The case of a constant $\tilde{\sigma}=\tilde{\sigma}_0$ has been already shown in (Calogero \& Velten,  2013). The observational data points displayed in Fig. 1 are based on a technique which uses the differential age of old red galaxies. They were compiled in (Farooq et al 2013). The main conclusion that can be drawn from the pictures is that the diffusion dynamics can be made arbitrarily close to those of the $\Lambda$CDM model by choosing the exponent $k$ positive and large, or the exponents $\lambda, \delta, h$ negative and with large absolute value.

\begin{figure}
\begin{center}
\subfigure[$\tilde{\sigma}_{(n)}$]{\includegraphics[width=0.45\textwidth]{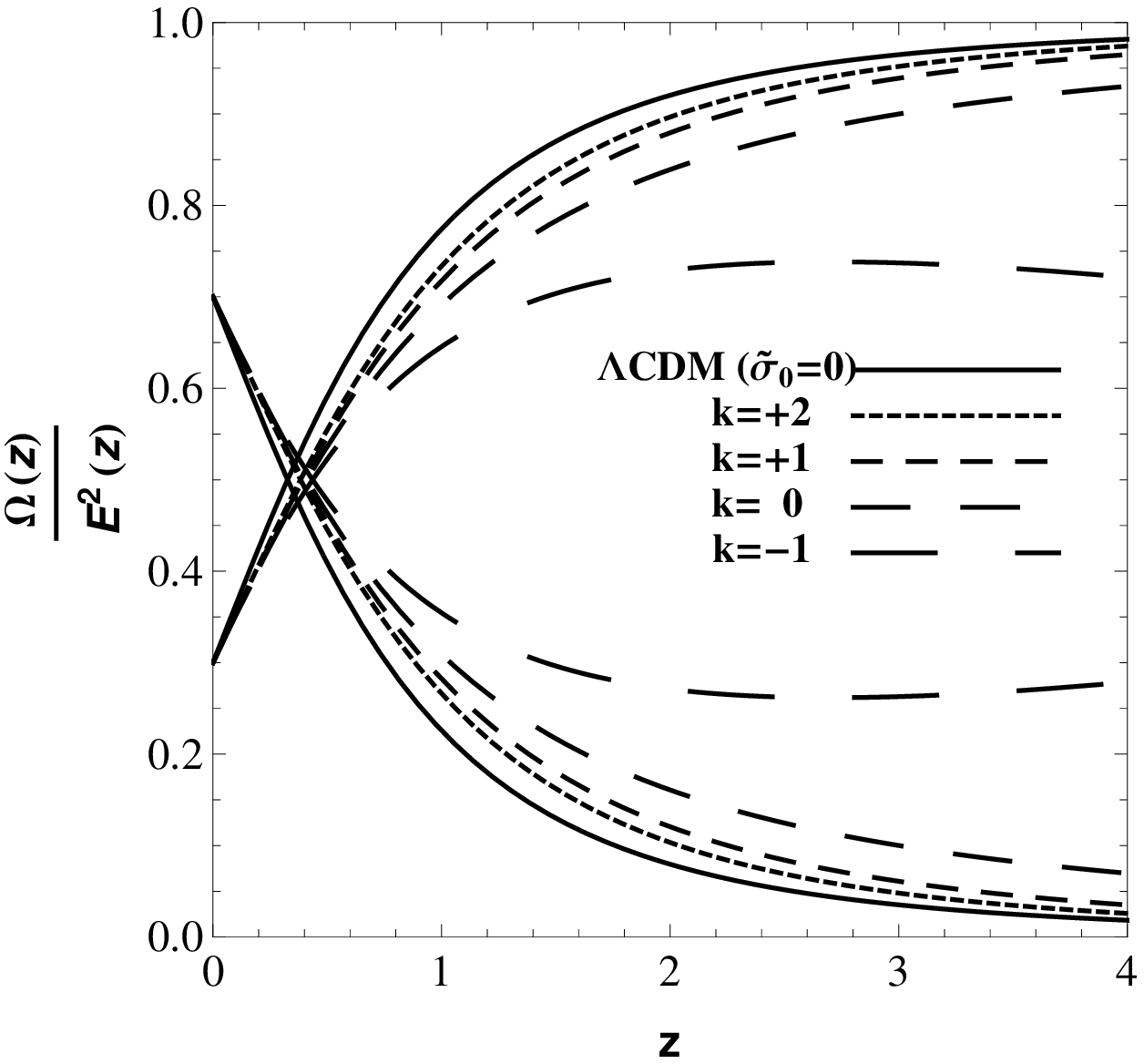}}\quad
\subfigure[$\tilde{\sigma}_{(\rho)}$]{\includegraphics[width=0.45\textwidth]{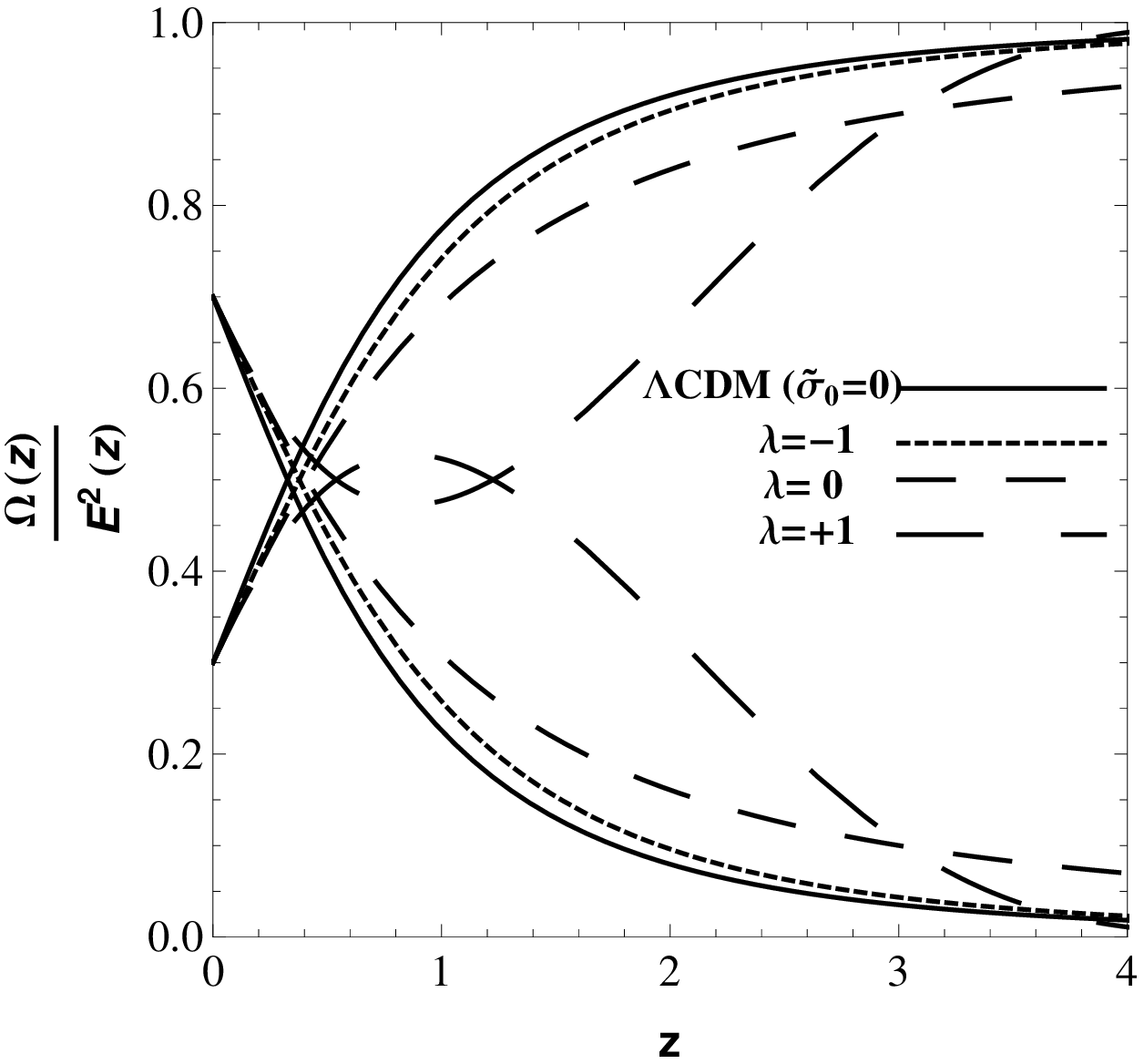}}\\
\subfigure[$\tilde{\sigma}_{(\phi)}$]{\includegraphics[width=0.45\textwidth]{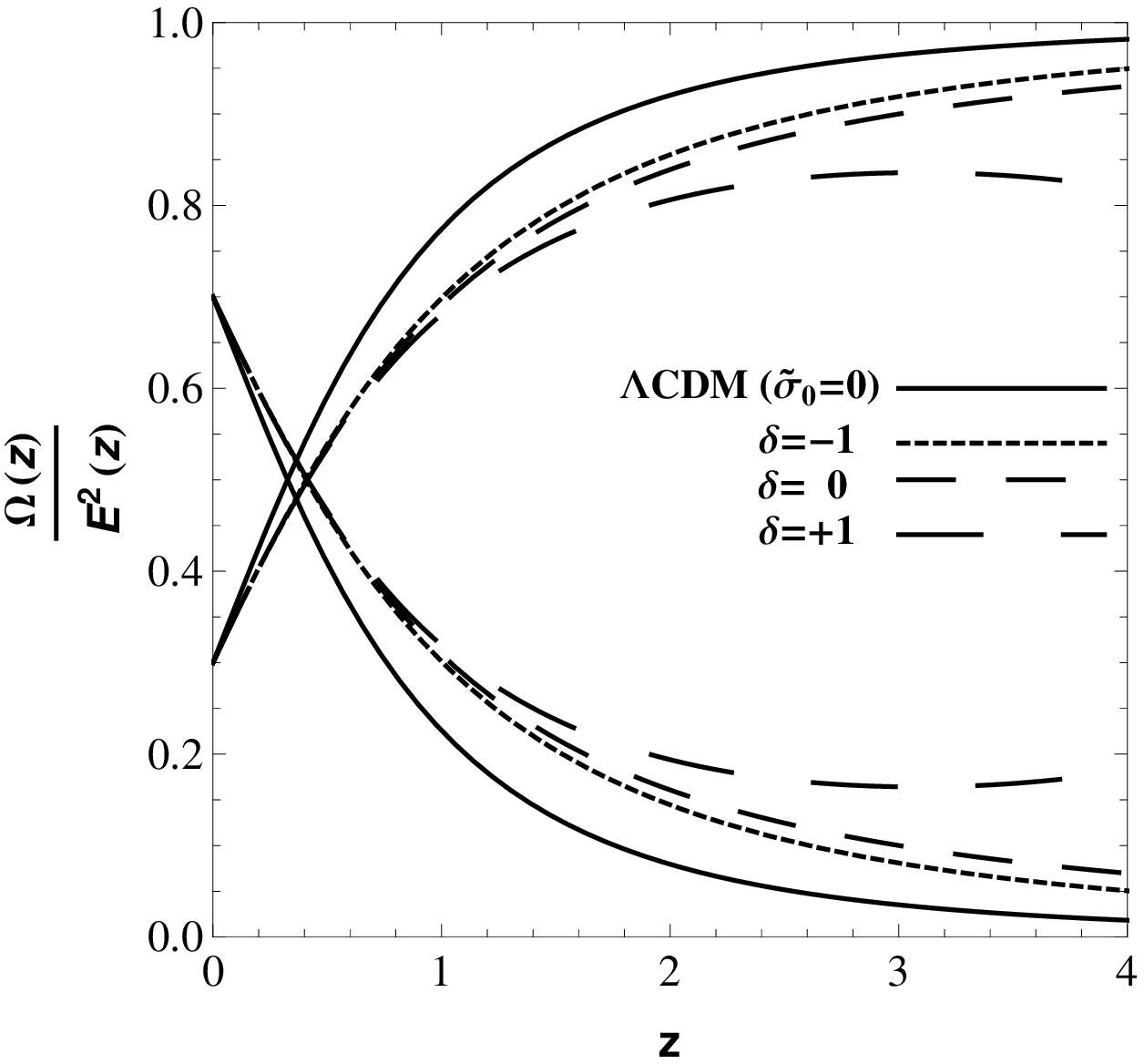}}\quad
\subfigure[$\tilde{\sigma}_{(H)}$]{\includegraphics[width=0.45\textwidth]{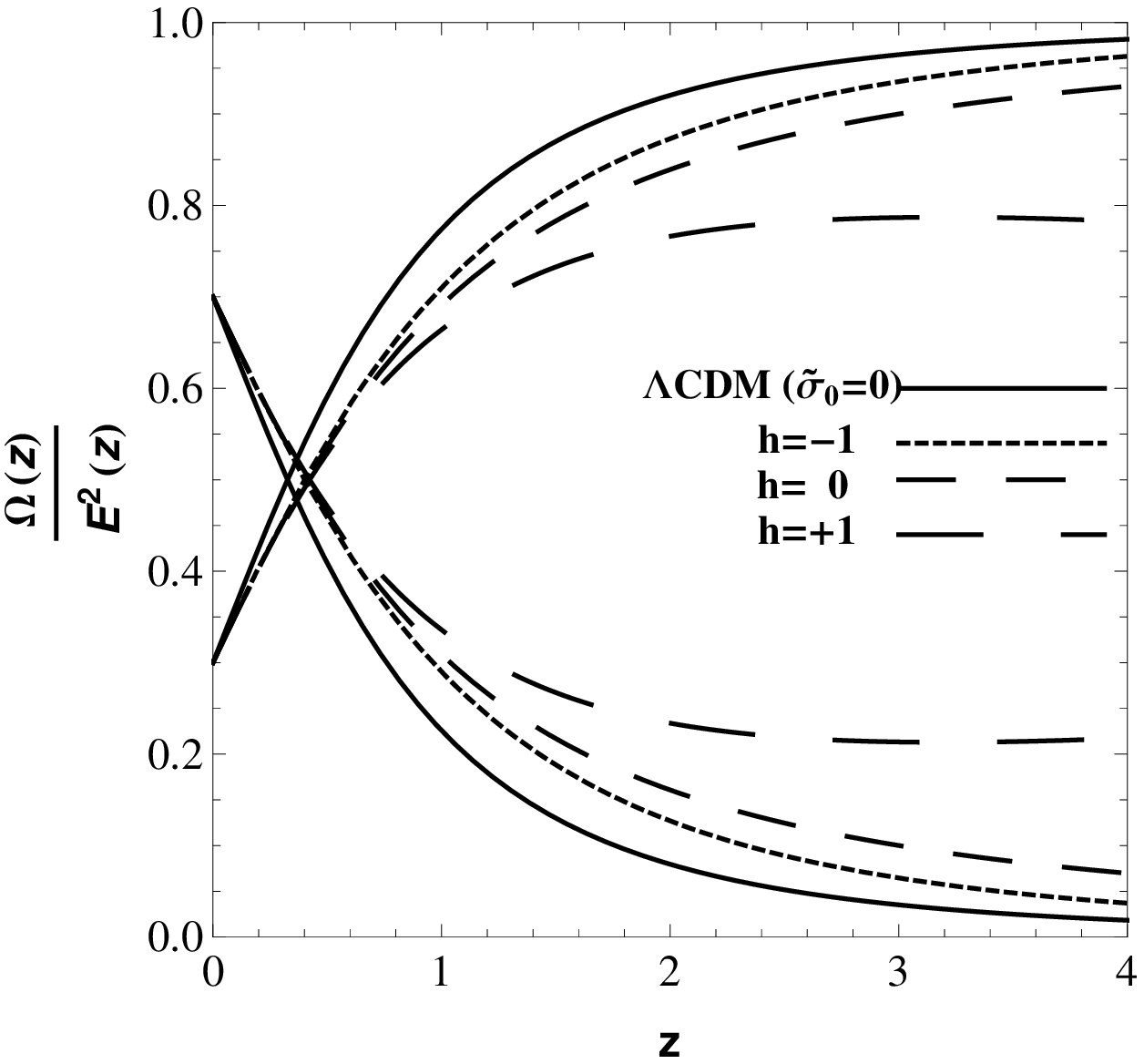}}
\caption{Evolution of the fractionary density parameters for the various choices of time dependent diffusion coefficient.}
\label{fig2}
\end{center}
\end{figure}

\section{Conclusions}

We have investigated the background evolution for a cosmological model where the matter component undergoes microscopic velocity diffusion into the dark energy field, which acts as the diffusion solvent.
Previously, the case of a constant diffusion coefficient $\sigma=const$ was studied~(Calogero \& Velten, 2013). In this contribution we consider different temporal dependences for the diffusion coefficient, which derive by postulating a power-law dependence of $\sigma$ on the other dynamical variables of the model.

Our main result can be stated in the following way:  By a proper choice of the exponent in the power-law, the dynamics of the diffusion model can be made arbitrarily close to those of the $\Lambda$CDM expansion. In some sense, this means that even for "high" values for the magnitude of the today's diffusion coefficient $\tilde{\sigma}_0$, such as $\tilde{\sigma}_0=0.1$, an appropriate time dependence can alleviate the diffusion effects on the cosmic background dynamics. In any case, following the results of~(Calogero \& Velten, 2013), it is mandatory a study making use of the cosmological perturbation theory. The cosmic matter diffusion is very sensitive to this analysis and the most strong constraints come from the structure formation process. We hope to deal with this issue in a future communication.

\textbf{Acknowledgement}:  HV thanks CNPq (Brazil) for partial financial support and the organizers of the II GRACO held in Buenos Aires.

\end{document}